\documentclass[a4paper,twoside]{article}
\newcommand{\affil}[1]{$^{\rm #1}$}
\usepackage[authoryear]{natbib}
\bibpunct{(}{)}{;}{a}{}{,}
\usepackage{graphicx}
\usepackage{psfig}
\usepackage{aas_macros}
\usepackage{epsf}
\usepackage{graphics}
\date{} 
\newlength{\plotwidth}
\newlength{\fullwidth}
\newcommand{\bj}{\mbox{$b_{\rm\scriptscriptstyle J}$}}
\newcommand{\rf}{\mbox{$r_{\rm\scriptscriptstyle F}$}}

\newcommand{\kms}{\mbox{\,km\,s$^{-1}$}}

\newcommand{\simlt}{\mbox{\lower.8ex\hbox{$\; \buildrel < \over \sim \;$}}}
\newcommand{\simgt}{\mbox{\lower.8ex\hbox{$\; \buildrel > \over \sim \;$}}}

\newcommand{\totalsixdf}{179\,262}    
\newcommand{\totalDB}{89\,211}     
\newcommand{\uniqueDB}{83\,014}     
\newcommand{\totalDBqual}{76\,443}           
\newcommand{\uniqueDBqual}{71\,627}           
\newcommand{\aimsixdf}{150\,000}

\newcommand{\fields}{936}

\newcommand{\kt}{\mbox{$K$}}

\newcommand{\plotone}[1]
    {\centering \leavevmode \vspace{-10mm} \psfig{file=#1,width=80mm,clip=} \vspace{-10mm}}

\newcommand{\plotfull}[2]
    {\centering \leavevmode \psfig{file=#1,width=#2\fullwidth,clip=}}
\newcommand{\plotrot}[3]
    {\centering \leavevmode \psfig{file=#1,width=#2\fullwidth,angle=#3,clip=}}

\title{\LARGE\bf\flushleft Second Data Release of the 6dF Galaxy Survey}
\author{\parbox{\textwidth}{\flushleft
\vspace{-0.5cm}
{\it D.~Heath Jones\affil{A,D}, 
Will Saunders\affil{B},
Mike Read\affil{C},
Matthew Colless\affil{B}
}\\
\vspace{0.4cm}
{\small \affil{A}\,Research School of Astronomy \& Astrophysics, The Australian
National University, Weston Creek, ACT 2611, Australia}\\
{\small \affil{B}\,Anglo-Australian Observatory, P.O.\ Box 296, Epping, NSW 2121,
Australia}\\
{\small \affil{C}\,Institute for Astronomy, Royal Observatory, Blackford Hill,
Edinburgh, EH9~3HJ, United Kingdom}\\
{\small \affil{D}\,\,Email: heath@mso.anu.edu.au}
}}

\begin{document}
\twocolumn[
\begin{changemargin}{.8cm}{.5cm}
\begin{minipage}{.9\textwidth}
\vspace{-1cm}
\maketitle
%
%
\small{\bf Abstract:}
The 6dF Galaxy Survey is measuring around 150\,000 redshifts 
and 15\,000 peculiar velocities from galaxies over the
southern sky at $|\,b\,|>10^\circ$. When complete, it will be the largest survey of its
kind by more than an order of magnitude. Here we describe the
characteristics of the Second Incremental Data Release (DR2) and
provide an update of the survey. This follows earlier data made public
in December 2002 and March 2004. A total of \uniqueDB\ sources now have
their spectra, redshifts, near-infrared and optical photometry available online
and searchable through an {\tt SQL} database at 
http://www-wfau.roe.ac.uk/6dFGS/.


\medskip
\medskip
\end{minipage}
\end{changemargin}
]
\small


\section{Survey Overview}

The 6dF Galaxy Survey\footnote{6dFGS home: http://www.aao.gov.au/local/www/6df/}
\citep[6dFGS;][]{jones04su}
is a combined redshift and 
peculiar velocity survey encompassing all but the most heavily obscured 
regions of the southern sky. The redshift survey aims to secure 
\aimsixdf\ redshifts across apparent magnitude limited samples complete to
$ (K, H, J, r_F, b_J) = $
(12.75, 13.05, 13.75, 15.60, 16.75).
In addition to these, a number of smaller subsamples 
defined from optical, near-infrared and radio
surveys, fill-out fibre allocations the sky over.
The peculiar velocity survey is taking the brightest 15\,000
E/S0 galaxies and combining their 6dF velocity dispersion measurements with
photometric parameters from Two Micron All-Sky Survey
(2MASS) near infrared imaging \citep{jarrett00} to yield Fundamental
Plane distances and peculiar velocities.

The 6dFGS data are made public at approximately yearly intervals, 
with Early and First Data Releases having taken place in December 2002 
and March 2004 respectively. Released redshift data and associated
photometric input
catalogues are searchable through an online 
database\footnote{Online database: http://www-wfau.roe.ac.uk/6dFGS/}
maintained by the Royal Observatory Edinburgh.
In this paper we present the 6dFGS data we are making available for 
the Second Incremental Data Release (DR2) of the survey
and summarise its main attributes. 

The wide sky coverage and near-infrared target selection of the 6dFGS
set it apart from other redshift and peculiar velocity surveys. Near-infrared
selection is most sensitive to the peak of the galaxy spectral energy 
distribution, which is dominated by old stars that make up most
of the stellar mass in the majority of galaxies. Compared to
optically-selected samples, which are biased towards younger,
star-forming galaxies, the near-infrared selection of the 6dFGS favours
older, bulge-dominated galaxies that are the 
best targets for Fundamental Plane distance determinations. Furthermore, near-infrared
selection is minimally affected by extinction due to dust, both internal
to the target galaxies and through the plane of our own Galaxy.
As a consequence, the orientation of target galaxies is not important,
and the 6dF survey fields can cover the entire southern sky apart from a
$|\,b\,|\leq10^\circ$ strip around the Galactic plane.

\begin{figure}
\plotone{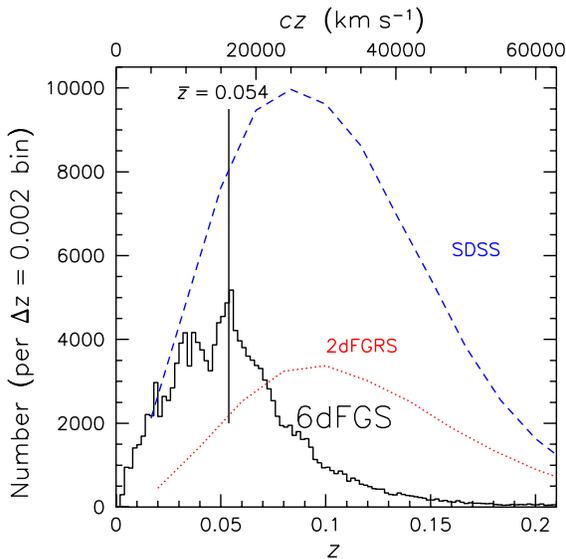}
\caption{Number-redshift distribution for the 6dF Galaxy Survey,
comprised of First Data Release galaxies with redshift quality 
$Q \ge 3$ and $cz > 600$~\kms.
The mean redshift for the survey ($\bar{z} = 0.054$)
is indicated with a vertical solid line.
Also shown are the equivalent distributions for the
SDSS ({\em short dashed line,blue}) and 2dFGRS ({\em dotted line, red}).
The 6dFGS and SDSS distributions have been normalised to yield
the final sample size expected for each, since neither survey is
yet to finish. 
}
\label{fig:nz}
\end{figure}

Figure~\ref{fig:nz} shows the redshift coverage of the 6dFGS compared
to the 2dF Galaxy Redshift Survey \citep[2dFGRS;][]{colless01} and the
Sloan Digital Sky Survey \citep[SDSS;][]{blanton01,york00}.
In terms of sky coverage, the 6dF Galaxy Survey will ultimately 
cover 8 times the area of 2dFGRS
and twice that of SDSS.
In terms of sample size, the 6dFGS will yield two-thirds as many galaxy redshifts as 
2dFGRS and one-fifth as many as SDSS.
Both of these deeper surveys have median redshifts double that of the median $\bar{z} \approx 0.05$
for 6dFGS. While these surveys have greatly advanced our knowledge of
the local universe, many questions are more readily addressed by
a survey such as 6dF that yields both galaxy mass and velocity together.

\begin{table*}
\begin{center}
\caption{The 6dFGS target samples used to define the tiling. There are
also samples of 6843 SUMSS sources (Sadler, Sydney; ${\rm id}=125$), 
466 Durham/UKST Galaxy Survey sources 
(Shanks, Durham; ${\rm id}=78$), and 724 Horologium Supercluster
sources (Rose et al, North Carolina) not used in the
tiling but included for serendipitous observation. A further
4096 orphaned sources were in the original target catalogues but are no
longer, due changes in the 2MASS source catalogues. However, their redshift
information has been retained under ${\rm id}=999$. Surveys with
larger weights have higher priority in the allocation of
fields. `Coverage' is as expected from the tiling simulations; in
practice fibre breakages and imperfect fibre assignment reduce these
numbers, especially for lower priority samples.
\label{tab:6dFGStargets}} 
\vspace{6pt}
\begin{tabular}{llccrc}
\hline 
id & Sample (Contact, Institution)       & Weight &  Total & Coverage \\
\hline
1& 2MASS $K_s<12.75$ (Jarrett, IPAC)       & 8     & 113988 & 94.1\% \\
3& 2MASS $H  <13.05$ (Jarrett, IPAC)   & 6     &   3282 & 91.8\% \\
4& 2MASS $J  <13.75$ (Jarrett, IPAC) & 6     &   2008 & 92.7\% \\
7& SuperCOSMOS $r_F<15.6$ (Read, ROE)   & 6     &   9199 & 94.9\% \\
8& SuperCOSMOS $b_J<16.75$ (Read, ROE)   & 6     &   9749 & 93.8\% \\
90& Shapley             (Proust, Paris-Meudon)   & 6     &    939 & 85.7\% \\
113& ROSAT All-Sky Survey  (Croom, AAO)  & 6     &   2913 & 91.7\% \\
119& HIPASS ($>4\sigma$) (Drinkwater, Queensland)    & 6     &    821 & 85.5\% \\
126& IRAS FSC $6\sigma$ (Saunders, AAO) & 6     &  10707 & 94.9\% \\
5& DENIS $J<14.00$    (Mamon, IAP)   & 5     &   1505 & 93.2\% \\
6& DENIS $I<14.85$    (Mamon, IAP)  & 5     &   2017 & 61.7\% \\
116& 2MASS AGN          (Nelson, IPAC)  & 4     &   2132 & 91.7\% \\
129& Hamburg-ESO Survey  (Witowski,Potsdam) & 4     &   3539 & 90.6\% \\
130& NRAO-VLA Sky Survey (Gregg, UCDavis)   & 4     &   4334 & 87.6\% \\
                      &          &               &     \\
Total                  &       &  167133 & 93.3\% \\
\hline 
\end{tabular}
\end{center}
\end{table*}

Table~\ref{tab:6dFGStargets} is  from \citet{jones04su} and
summarises the breakdown of source
catalogues contributing to the master target list. Also shown is the
relative weighting (priority) given to each sample in the target allocation process, 
and the maximum coverage expected for each survey given that field 
allocation. These coverage fractions are theoretical values which will be
reduced by fibre-breakage rates in practice, particularly for lower
priority programmes. In total there are
167\,133 objects allocated, of which two-thirds are 
in the near-infrared-selected sample.

\citet{jones04su} describe the main facets of the 6dF Galaxy Survey,
including the Six-Degree Field instrument, target catalogue
construction, and the allocation of fields and fibres to targets.
The data reduction and redshifting procedures are also described,
along with the main characteristics of the data comprising the
First Data Release. The interested reader is referred to this paper
for detailed information about the 6dFGS. \citet{campbell04}
describe the algorithm used to optimise the placement of field centres 
with respect to the galaxy distribution on the sky. A series of 
{\sl AAO Newsletter} articles trace the history of the 6dF instrument 
\citep{watson98,watson99,saunders01}, 
target selections \citep{saunders03,mauch02}, 
online database \citep{read03} and First Data Release \citep{saunders04}.
Various conference proceedings give further information about the 
instrument \citep{parker98,watson00}, survey aims \citep{wakamatsu03,
colless03} and First Data Release \citep{jones05}.
All of these references can be found together on the 6dFGS
Publications Page\footnote{Publications:\\
http://www.aao.gov.au/local/www/6df/publications.html}.

The 6dFGS \kt-band luminosity function \citep[LF;][Jones et~al., in prep]{jones05}, shows
excellent agreement with earlier work using 2MASS with 2dFGRS
\citep{cole01}, ZCAT \citep{kochanek01} and SDSS \citep{bell03}.
Considering the precision of these modern LFs and the influence
of such factors as sky coverage and nearby large-scale structures, the
level of agreement is remarkable. Other near-infrared and optical luminosity functions
from 6dFGS for \bj\rf$JH$ (\citeauthor{jones05})
also show good agreement.

The 6dFGS is being utilised for a variety of other programmes, including:
\begin{itemize}
\item Study of the AGN and QSO populations from 2MASS (Francis et al., ANU; 
        Cutri et al., Caltech).
\item Radio-selected samples (Mauch and Sadler, Sydney; Mamon, IAP;
        Doyle and Drinkwater, Queensland).
\item Galaxy ages and metallicities (Proctor and Forbes, Swinburne; Lah, ANU;
Colless, AAO).
\item Southern sky peculiar velocity survey (Campbell and Colless, AAO; Lucey, Durham),
\item Galaxy groups, nearby superclusters and voids (Brough, Kilborn and Forbes, Swinburne; 
Wakamatsu, Gifu; Johnston-Hollitt et al., Tasmania).
\item Large-scale structures and clustering (Lahav and Rassat, UCL; Fairall et al., Cape Town).
\end{itemize}

In Section~2 we summarise the scope and features of the Second Incremental
Data Release for the 6dFGS. Section 3 outlines caveats and cautions that users
should bear in mind when using the data. Comments regarding the
survey timeline from now until completion are given Section 4.

\begin{figure*}
\plotfull{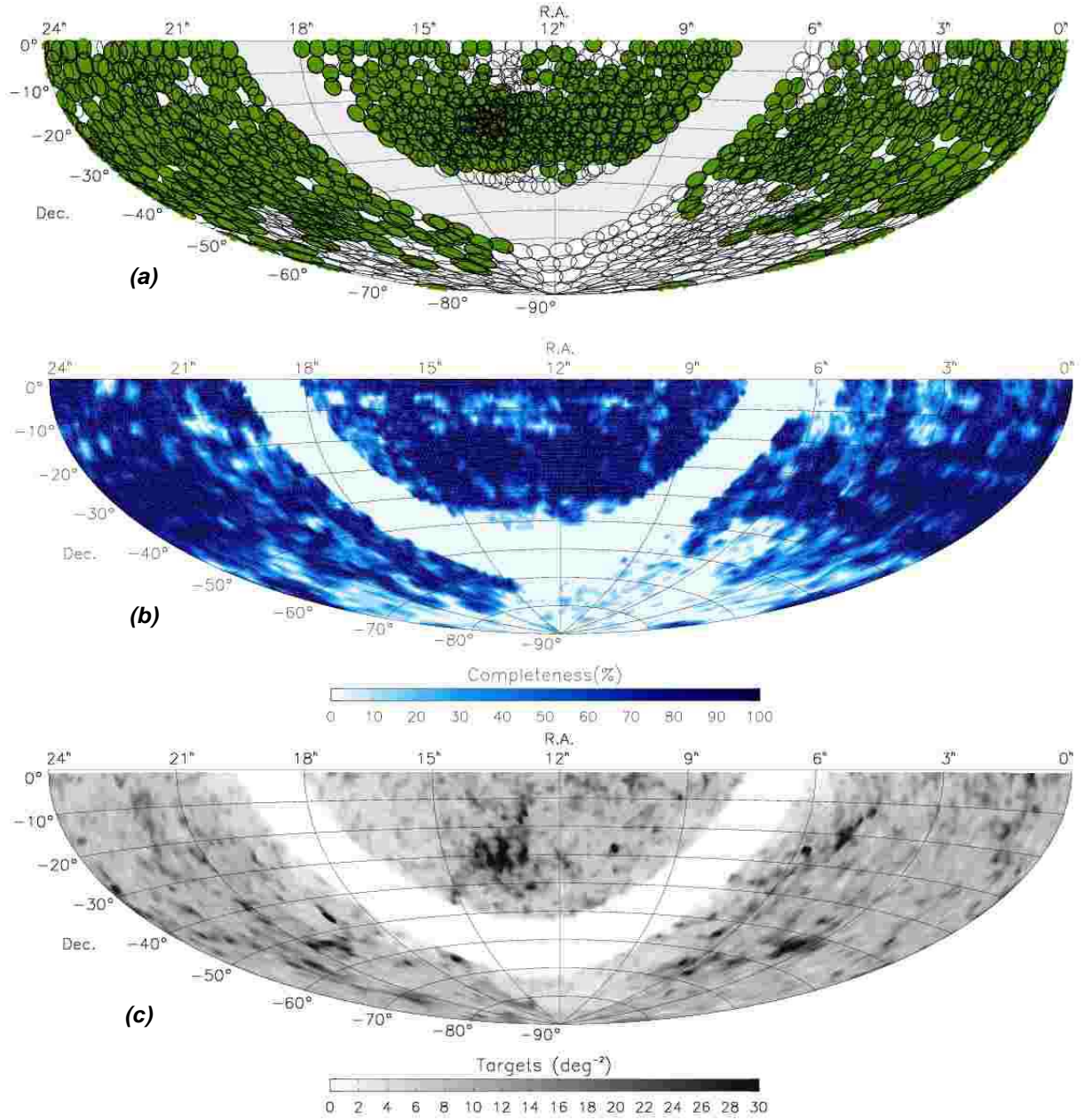}{1.0}
\caption{($a$) Location of the observed fields ({\em solid discs}) 
contributing redshifts to the First Data Release. The tiling pattern of all fields 
({\em open discs}) is also shown.
($b$) Redshift  completeness on the sky, combining the 6dF First Data Release redshifts 
with the literature sources.
($c$) Density of targets on the sky.
  }
\label{fig:obsflds}
\end{figure*}

\section{Second Data Release}

\subsection{Scope of DR2}

The Second Incremental Data Release (DR2) for the 6dF Galaxy Survey
spans observations during the period January 2002 to October 2004,
including and superceding DR1. It contains \totalDB\ spectra that have yielded \uniqueDB\ unique 
galaxy redshifts over roughly two-thirds of the southern sky.
Within these sets, the number of database spectra with acceptable quality 
redshift is \totalDBqual, where `acceptable' means quality parameter
$Q  = 3$ or 4 \citep[see ][]{jones04su}. 
The number of $Q = 3, 4$ unique redshifts
is \uniqueDBqual. As discussed below, only redshifts with $Q=3$ or 4 should be 
used for extragalactic science.

Extragalactic redshifts are ranked on a quality scale of $Q = 1$ to 4 by
human operators, according to the definitions given in Sec.~4.4 of \citet{jones04su}.
$Q=1$ represents spectra of no value, $Q=2$ for tentative redshift
values, $Q=3$ for probable redshifts and $Q=4$ for reliable redshifts.
Only $Q=3$ or 4 redshifts should be used for astrophysical
applications, although some QSOs classified earlier in the survey will
have a $Q=2$ because no QSO template existed at the time to confirm the redshift
Recently a new category, $Q=6$, was introduced and this is discussed in Sect.~\ref{redshifts}.

DR2 takes its data from  \fields\ fields. Figure~\ref{fig:obsflds}($a$) shows that their 
distribution is true to observing strategy adopted by 6dFGS from the outset: mid-latitude fields
were tackled first, followed by those nearest the equator and finished with the polar 
targets. Field coverage along
the central declination strip $-42^\circ < \delta < -23^\circ$ is largely complete,
as is coverage of the $-23^\circ < \delta < 0^\circ$ equatorial band, save for regions
around 11 -- 13\,hr and 03 -- 06\,hr right ascension. The former is a region already
covered extensively by the SGP region of the 2dF Galaxy Redshift Survey \citep{colless01}.
Substantial progress has also been made on some
of the $\delta < -42^\circ$ polar fields, which have been included in DR2.

Figure~\ref{fig:obsflds}($b$) shows the distribution of galaxy redshift
completeness on the sky. This is the fraction of galaxies in the original target
catalogue with acceptable ($Q = 3$ or 4) extragalactic redshifts, either new from 
6dFGS or existing literature ones. Variations in completeness at this stage in the
survey are dominated by variable observing conditions, since it is only in the final
stages of the survey that virgin fields become sparse enough to allow
re-observing fields with low completeness. Figure~\ref{fig:obsflds}($c$) shows
the density of 6dFGS targets on the sky. The variation of source density is a major
challenge. The strategies
devised to handle this are described elsewhere \citep{jones04su, campbell04}.
 
\begin{figure}
\plotone{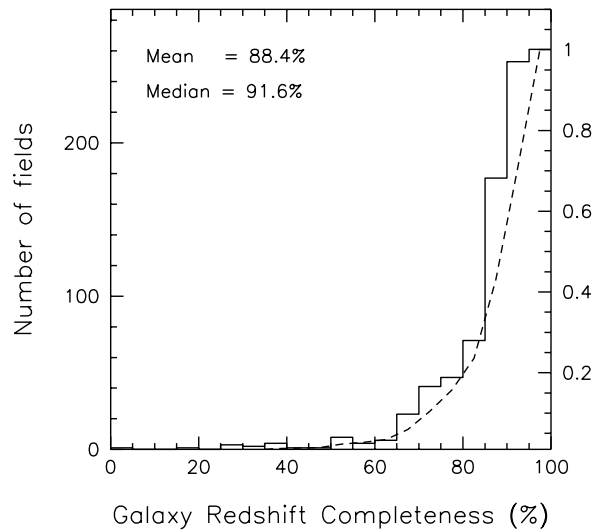}
\caption{Galaxy redshift completeness by field, where completeness
  is the number of 6dF redshifts over the total 6dF redshifts and
  failures. The dashed line indicates the cumulative fraction according to 
  the right-hand axis.}
\label{fig:fldcompl}
\end{figure}

Another measure of survey efficiency is field completeness, shown in
Fig.~\ref{fig:fldcompl}. The field completeness is the number of 
$Q = 3$ or 4 extragalactic redshifts from all targets in a field, and as
such, only applies to redshifts from 6dF. This shows little change from
DR1: about 80 per cent of all fields are 80 per cent complete or more,
and more than half have completeness in excess of 85 per cent.

\begin{figure*}
\plotrot{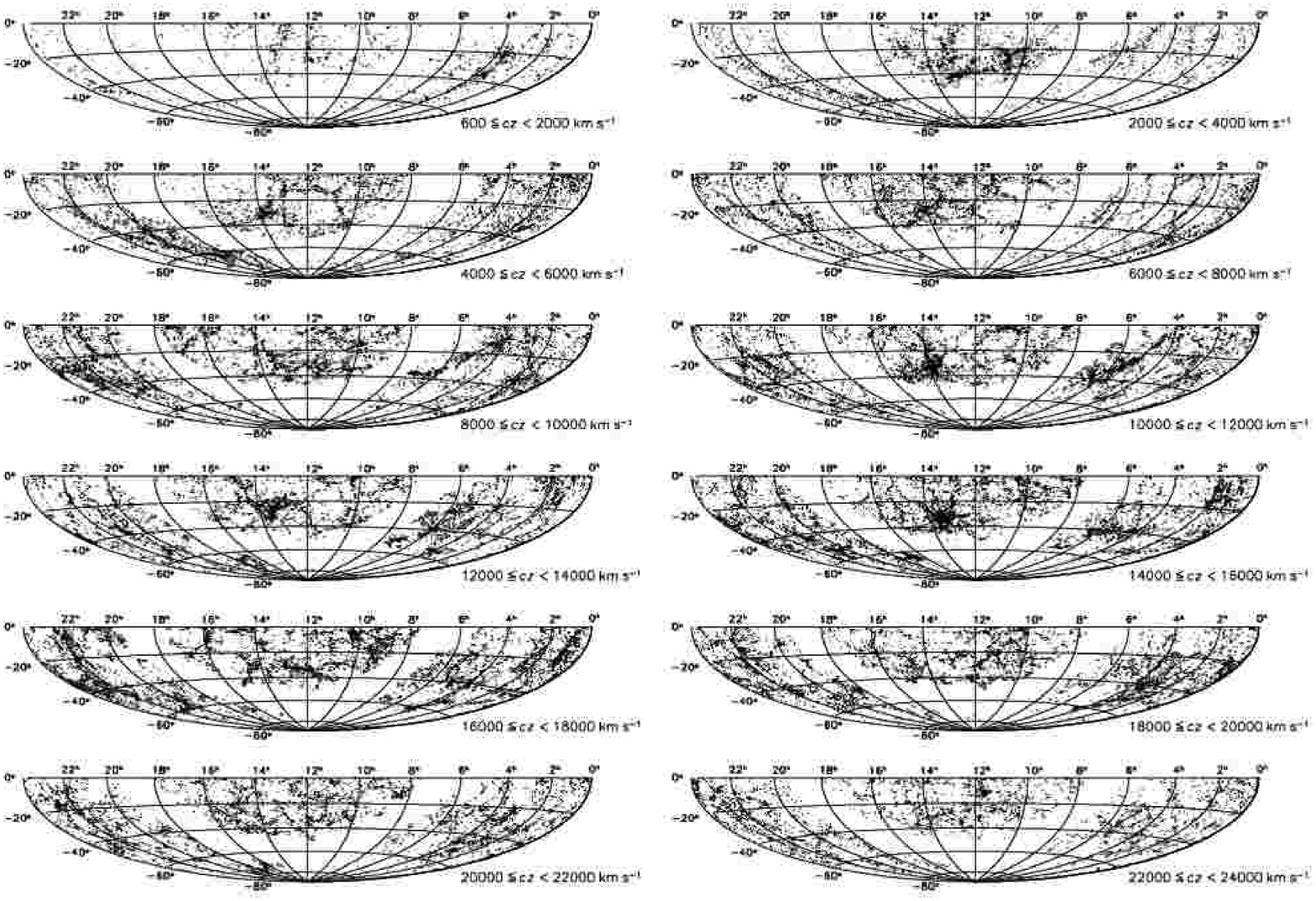}{1.0}{90} 
\caption{All-southern-sky projections of the 6dF Galaxy Survey to date. Each
panel corresponds to a 2000\kms-wide hemispherical shell in 
recession velocity. Polar coverage is still largely incomplete.}
\label{fig:mulait0}
\end{figure*}

\begin{figure*}
\plotrot{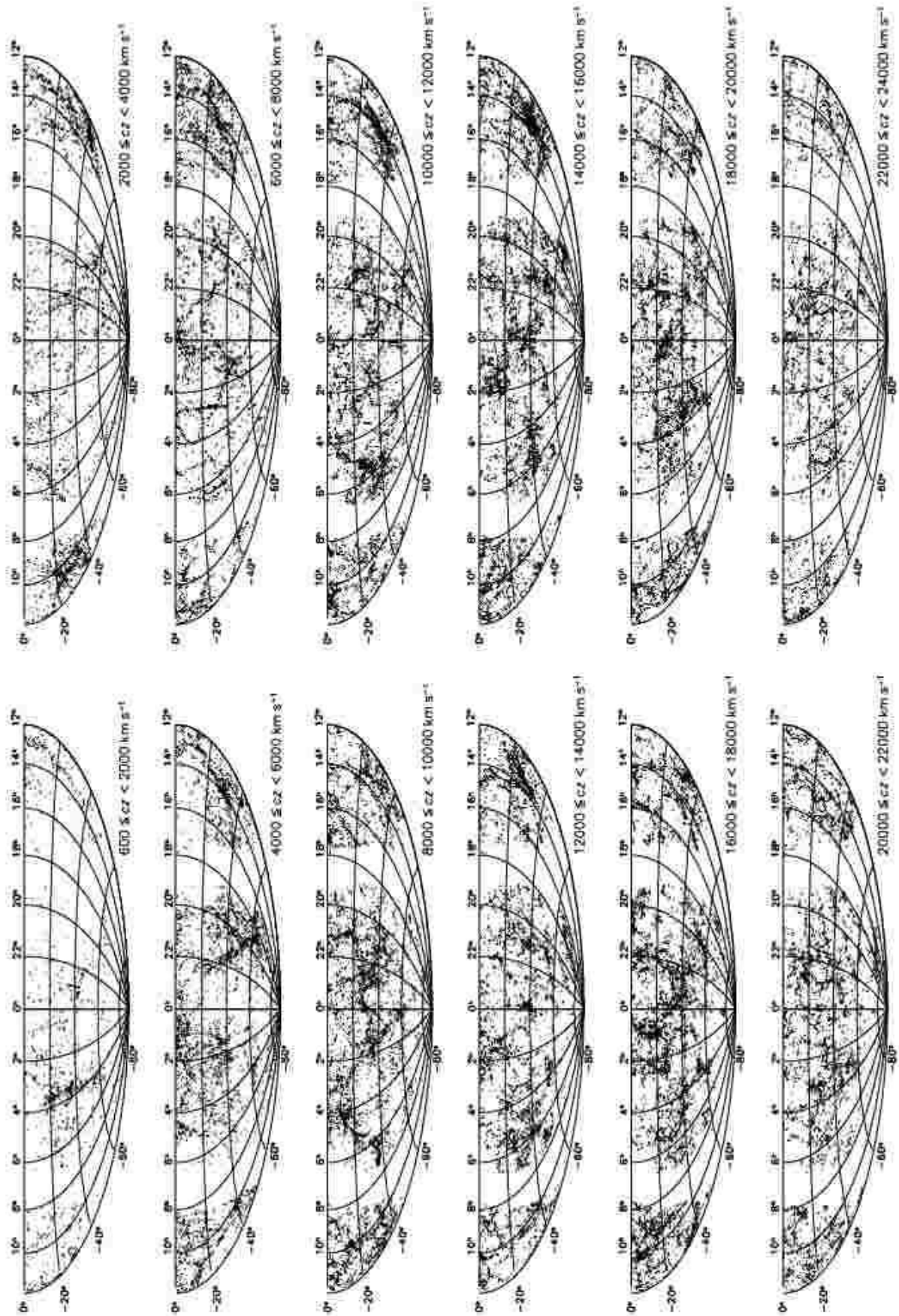}{1.0}{90} 
\caption{Same as Fig.~\ref{fig:mulait0} but rotated through 12 hours in
right ascension.}
\label{fig:mulait180}
\end{figure*}

Figures~\ref{fig:mulait0} and \ref{fig:mulait180} show the distribution of DR2 redshifts
on the sky. They show many local large-scale structures, such as the Shapley
and Hydra-Centaurus Superclusters; this is the most detailed and
comprehensive view of the southern local universe to date.

\subsection{Online Database}

The 6dF Galaxy Survey Online Database\footnote{Online database:
http://www-wfau.roe.ac.uk/6dFGS/ }
is maintained by the Wide Field Astronomy
Unit of the Institute for Astronomy, University of Edinburgh.
The site has been operating since an early
6dF data release in 2002 December of 17\,000 redshifts. This was followed by
the First Incremental Data Release of 6dFGS data in March 2004 (DR1), which increased
the online catalogue to 52\,048 measured spectra. This paper marks the Second
Data Release; a third and final release will be made available on
completion of the survey.

The database is based on the format of the 2dFGRS database and uses
Microsoft's {\tt SQL}~{\sl Server 2000}.  Each survey object has a corresponding multi-extension
FITS file containing postage-stamp images of the object in \bj\rf$JHK$ and its
6dF-measured spectrum. Structured Query Language ({\tt SQL}) is used to interrogate the
database and extract its contents. 

Fuller descriptions of the database can be found in Sec.~5.2
of \citet{jones04su} and \citet{read03}. What follows here is a summary of its
main features.

\begin{table*}
\begin{center}
\caption{Tables of data in the 6dFGS Database 
\label{tab:database1}
} 
\vspace{6pt}
\begin{tabular}{llc}
\hline \hline
Table name   &        Description  &   Programme  \\
             &                     &   ID Numbers  \\

\hline

{\tt TARGET}    &   the master target list                  &{\tt progid} \\
{\tt SPECTRA}   &   redshifts and observational data               &  $-$    \\
{\tt TWOMASS}   &   2MASS input catalogue $K$, $H$, and $J$        &  1, 3, 4\\
{\tt SUPERCOS}  &   SuperCOSMOS bright galaxies $b_J$ and $r_F$    &  7, 8   \\
{\tt FSC}       &   sources from the IRAS FAINT Source Catalogue   &  126    \\
{\tt RASS}      &   candidate AGN from the ROSAT All-Sky Survey    &  113    \\
{\tt HIPASS}    &   sources from the HIPASS HI survey              &  119    \\
{\tt DURUKST}   &   extension to Durham/UKST galaxy survey         &  78     \\
{\tt SHAPLEY}   &   galaxies from the Shapley supercluster         &  90     \\
{\tt DENISI}    &   galaxies from DENIS $I < 14.85$                &  6      \\
{\tt DENISJ}    &   galaxies from DENIS $J < 13.85$                &  5      \\
{\tt AGN2MASS}  &   candidate AGN from the 2MASS red AGN survey    &  116    \\
{\tt HES}       &   candidate QSOs from the Hamburg/ESO Survey     &  129    \\
{\tt NVSS}      &   candidate QSOs from NVSS                       &  130    \\
{\tt SUMSS}     &   radio source IDs from SUMSS and NVSS           &  125    \\

\hline \hline
\end{tabular}
\end{center}
\end{table*}

Data are grouped into several tables, listed in Table~\ref{tab:database1}. At its
heart is the {\tt TARGET} table, which contains the master target list used to
define 6dFGS observations. Every table is linked through the parameters
{\tt TARGETID} and {\tt TARGETNAME} which, although unique in the {\tt TARGET}
table, are not necessarily unique in the others (for example, some objects have been 
observed more than once). 
The 6dFGS parameterised spectral data comprising the redshift measurements and observational 
meta-data  are
held in the {\tt SPECTRA} table. The other tables constitute related source catalogues
(and their photometric data) that are used to define the various target subsets summarised in 
Table~\ref{tab:6dFGStargets}. 

In the {\tt TARGET} tables there are
4\,177 $Q=2$ sources, 5\,347 with $Q=3$ redshifts and 70\,391 with $Q=4$.
In the {\tt SPECTRA} table, however, there are multiple observations of the same
redshift: 7\,855 for $Q=2$, 6\,031 for $Q=3$ and 70\,047 for $Q=4$.
In addition, there are 4\,865 value-less $Q=1$ spectra which are kept only
for archival purposes and 292 recent Galactic sources ($Q=6$). There are
many more confirmed Galactic sources with $Q=2$ and noted with a comment.

{\tt TARGETNAME.FITS} is the name given to the multi-extension FITS file holding
all the data for a given target.
Postage-stamp FITS images are held in the first five extensions, from \bj\ and \rf\ 
SuperCOSMOS scans and $J$, $H$ and $K$ 2MASS imaging respectively.
As 6dF spectra are measured and then ingested into the database, this information
is held in FITS extensions six to eight, holding the V, R and VR-combined
spectra respectively.  Repeat observations (if any), are appended in the form of
further FITS extensions. \citet{jones04su} describe this arrangement in more detail.

The database can be queried in one of two ways. One web-based form
allows the user to make a number of basic selections from a number of common
operations. Another form permits users conversant in {\tt SQL} to construct
their own query directly. Users are also able to upload a list of objects
and cross-match this with the 6dFGS observations. Output is available as an HTML page returned directly
to the user, a downloadable file, and/or a TAR file of FITS data.
Several examples of database access are provided on the web site\footnote{
http://www-wfau.roe.ac.uk/6dFGS/examples}.

\subsection{Changes Since DR1}

The structure of the database and the data within are largely unchanged
since DR1. However, improved derivations for \bj\ and \rf\ galaxy
photometry have led to the introduction of two new magnitude parameters
in the {\tt TARGET} table: 
{\tt BMAGSEL} and {\tt RMAGSEL}. These hold the old \bj\ and \rf\ magnitudes
used to select the 6dFGS samples in these bands and   
formerly held by the {\tt BMAG} and {\tt RMAG} parameters.
The {\tt BMAG} and {\tt RMAG} parameters remain, although, the magnitudes they
now hold are from the improved \bj\ and \rf\ photometry introduced for the first 
time in DR2.

The new magnitudes result
from a significant improvement in the photometric calibration of the
photographic plate material. The original SuperCOSMOS magnitudes used to
select the optical samples suffer from scatter in the field-to-field
zeropoints of some tenths of a magnitude.  In 2003, J.~Peacock,
N.~Hambly and M.~Read re-calibrated all of the \bj, \rf\ (and 
$i_{\rm N}$) SuperCOSMOS magnitudes to a common zeropoint. This
recalibration work was originally done for the 2dFGRS and is described
on the 2dFGRS web site\footnote{2dFGRS photometric calibration:\\
{\tiny
http://www2.aao.gov.au/2dFGRS/Public/Release/PhotCat/photcalib.html}
}
The resulting zeropoint scatter has now been reduced to a few hundredths
of a magnitude (J.~Peacock, priv.\ comm.). Neither the old nor new
magnitudes have been corrected for Galactic extinction. The old
magnitudes must be used when considering the selection of the optical
samples, but the new magnitudes are otherwise preferred.

Pairing the recalibrated photographic magnitudes with the 6dFGS targets
is not a straightforward process, particularly for the brighter sources. 
A typical scenario will be that the target position will be paired with
several objects in a given SuperCOSMOS/UK Schmidt Telescope passband. For example,
these objects could be the parent object plus several daughter objects.
As the targets are bright and will often have structure it is likely that
the parent is the optimum object to get the magnitude for. In other cases
the parent will be a star/galaxy, galaxy/galaxy or star/star blend, and the
correct object is one of the daughters.

We have adopted an empirically-derived scheme that uses both proximity,
magnitude and blend parameter to assign identifications and classify
the match.  
Of the \totalsixdf\ target sources put through
the \bj-matching exercise,  
44 per cent had both proximity and magnitude rankings of 1. Of the remainder,
48 per cent had either a proximity or magnitude rank of 1, 
7 per cent had neither ranked as 1, and
1 per cent had no match at all. The corresponding numbers for \rf-matching 
were 
around 46, 46, 7 and 1  per cent.

\begin{figure*}
\plotfull{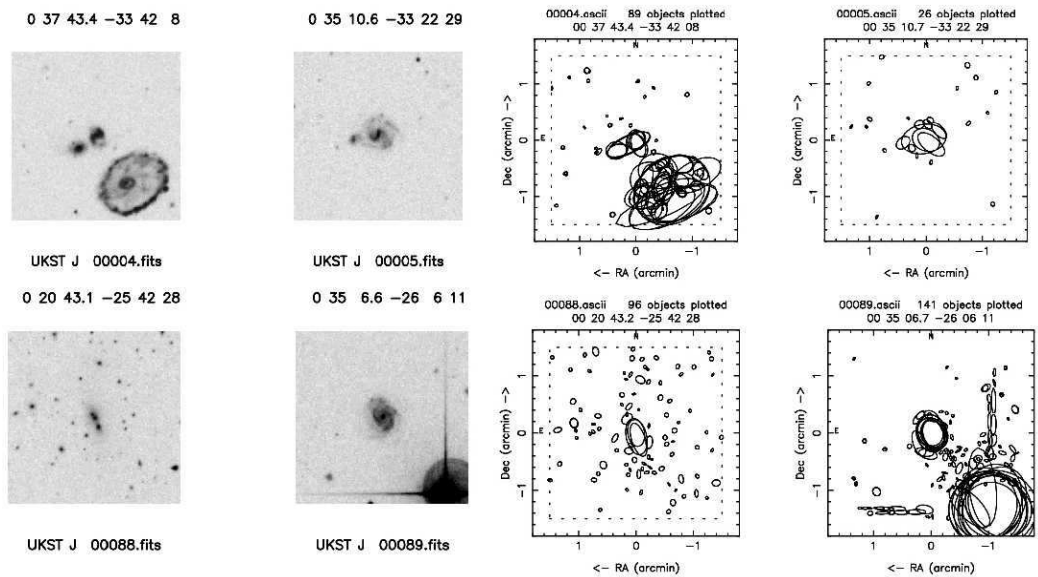}{0.9}
\caption{Examples of UK Schmidt telescope fields ({\em left}) that have resulted in poorly 
deblended  SuperCOSMOS targets ({\em right}). Included is one field encompassing the well-known
Cartwheel Galaxy, ESO\,350-\,G\,040 ({\em upper left panel}).}
\label{fig:deblend}
\end{figure*}

The non-matches fall into a few categories. Most are {\sl very} bright objects,
which in some cases have also been deblended into multiple targets by 2MASS.
Some are  $\bj \sim 14$  to 17 galaxies that lie near satellite trails.
Others are bright objects with poor deblending, globular clusters, 30 Doradus
in the LMC, and other extended sources; a few are blank bits of sky.
The null value assigned to {\tt BMAG} and {\tt RMAG} in the case 
of non-matches is 0.00. Figure~\ref{fig:deblend} shows examples
where the SuperCOSMOS deblending has led to incorrect 6dFGS target-matching.

Users of the 6dFGS database should take account of these limitations and
treat this preliminary matching with care. The matching algorithm will
be further refined for the final survey data release.

The only other changes to the database are improved descriptions
for some of the table parameters.


\section{Caveats and Limitations}

\subsection{Spectral Data}

With the 6dFGS still underway at the time of DR2, users should be
mindful that DR2 is an intermediate data release. Just as DR2 is a
significant improvement on DR1, so the quality, precision and
reliability of the survey data will be further refined before the final
data release at the completion of the survey.

Sec.~4.3 of \citet{jones04su} discusses in depth some of the typical
problems that can afflict the spectral data. Generally, the splicing of
V and R spectral data is acceptable, with the exception of a small
number of cases. The effects of strong fringing are occasionally
evident in the form of an oscillating spectral response. In some cases,
the quality of the V and R data are significantly different, usually
with the V being worse because of the lower counts in that
spectral range. While it is possible in principle to derive redshifts
from V or R halves alone, no half-data have been included in DR2.

Users should also be aware of spurious spectral features around
4440~\AA\ in V, and 6430 and 6470~\AA\ in R. These are due to
ghost reflections caused by the VPH gratings, which have been used
continuously from 2002 September onwards. These features are
typically $\sim 10$ pixels wide. 

Occasionally, poorly spliced data
will show a spurious feature at the spectral join around 5570~\AA.

Flux calibration of spectra remains highly approximate and  
users are discouraged from using the database for spectrophotometric
applications. Sec.~4.2 of \citet{jones04su} outlines the
spectral reduction steps and how they affect the final spectra.

Fibre cross-talk can occur when light from bright stars or Galactic
emission-line sources finds its way into parked fibres. The bright spectral
features contaminate adjacent spectra, causing spurious features.
We strongly advise that any peculiar object spectra discovered by
users are checked for this effect. To this end, we have made available 
the fully-reduced 2-d frames of spectra in the Downloads section of the
Online Database.

\subsection{Redshifts \label{redshifts}}

For data from 2004 August onwards, a new version of the redshifting
software was used that introduced some changes worth noting.
First, only sources with an initial software-assigned $Q$-value of 3 or less
were inspected by eye. While greatly increasing the efficiency of
redshifting the spectra, the original configuration of the software resulted in 
38 spurious high redshift sources that are in fact more likely misclassified low redshift
sources. For DR2, these sources have been flagged and re-assigned $Q=2$; they
will be redshifted again manually at a later date. The software is now
run in a different configuration that flags sources falling into
this category.

The second software change affecting post-August 2004 data is that
confirmed Galactic sources (stars, ISM Balmer-line emission, planetary
nebulae, and the like) are now assigned $Q=6$. Before this, they had been
given $Q=2$ and their nature noted in a separate comment. Either way,
all acceptable extragalactic redshifts have $Q=3$ or~4. The two
schemes will be unified for the final data release.

A final change to the software was the introduction of
a QSO template, which has dramatically improved the successful
identification of the higher redshift QSOs uncovered by the survey.

\begin{figure*}
\plotfull{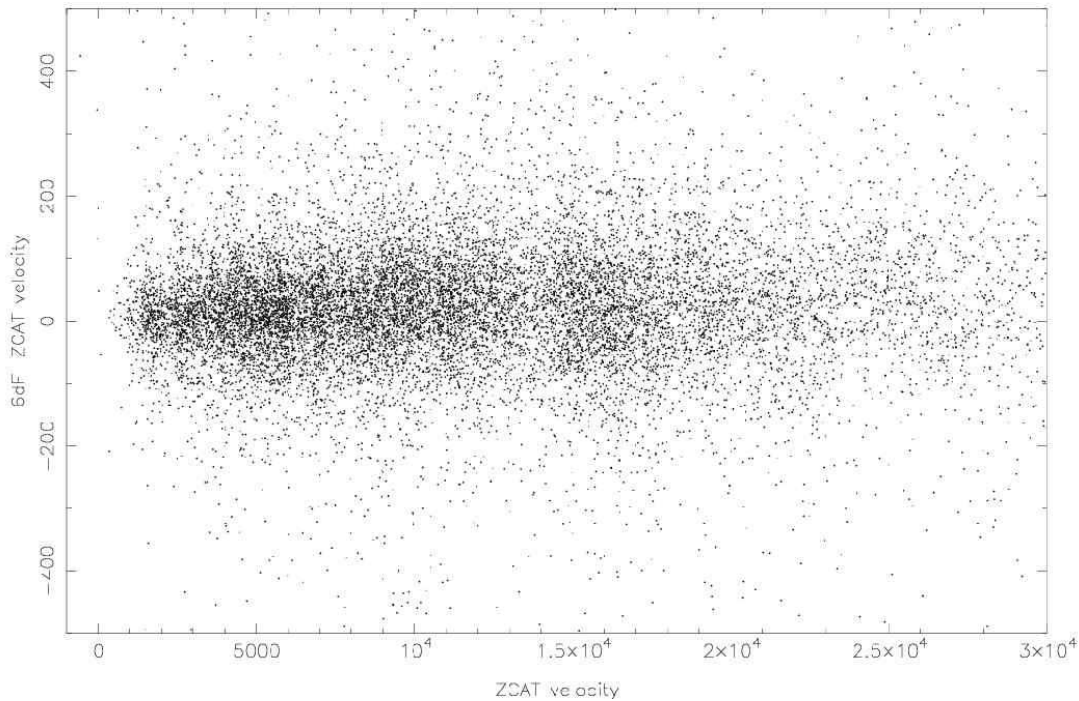}{0.9}
\caption{Redshift residuals $cz({\rm 6dFGS}) - cz ({\rm ZCAT})$ for a sample of 16\,127
sources with redshifts from 6dFGS and ZCAT. Both axes are in units of \kms.}
\label{fig:off}
\end{figure*}

To check the precision of the redshifts we compared a sample of 16\,127 6dFGS galaxies
with previous redshifts from Huchra's ZCAT Catalog (Fig.~\ref{fig:off}). 
The mean offset $cz({\rm 6dFGS}) - cz ({\rm ZCAT})$ was $22.0$\,\kms.
Repeat measurements indicate that most of the scatter is due to the ZCAT
sample, for which the source material is highly inhomogeneous.

\subsection{Photometry}

We have already discussed the limitations of the matching algorithm used
to pair the recalibrated \bj\ and \rf\ SuperCOSMOS magnitudes with 6dFGS
targets in instances where source deblending fails. The sources to watch
are those that are either bright ($\bj \simlt 15$), or have one or more
neighbours. Even for isolated targets, deblending may still be an issue
if the source is a bright spiral galaxy with genuine clumpy structure.
In general, users are urged to treat the optical photometry with
caution.


\section{Summary}

We have described the Second Incremental Data Release (DR2) for the 6dF
Galaxy Survey and its access through an online database. We have
discussed several issues affecting data quality and underscored the
preliminary and evolving nature of the data set. We urge potential users
of the data set to familiarise themselves with these caveats and with
the changes implemented since previous data releases.
 
Observing for the 6dFGS will nominally finish on 31 July 2005, although
some clean-up observations may occur after that date. A third and final
data release will be made sometime in 2006.

\section*{Acknowledgments} 

We are grateful to the staff of the Anglo-Australian Observatory,
who were responsible for the design and construction of the 6dF
instrument. AAO support for the 6dF Galaxy Survey continues
in the form of the observing and data reduction done by staff of the
UK Schmidt Telescope, without whom the project would not have been
possible. We also thank E.~Westra, M.~Williams, V.~Safouris,
and S.~Prior for their ongoing efforts in measuring redshifts from the survey. 
D.~H.~Jones is supported as a Research
Associate by Australian Research Council Discovery--Projects Grant
(DP-0208876), administered by the Australian National University.


\end{document}